\documentstyle[12pt]{article}
\newtheorem{Theorem}{Theorem}
\begin{document}
\begin{center}
\begin{large}
{\bf Variational Characterization of the Speed of Propagation of
\\ Fronts
for the Nonlinear Diffusion Equation}
\end{large}
\end{center}
\vspace{.1cm}
\begin{center}
R.\ D.\ Benguria and M.\ C.\ Depassier\\
        Facultad de F\'\i sica\\
        P. Universidad Cat\'olica de Chile\\
               Casilla 306, Santiago 22, Chile
\end{center}

\date\today
\begin{abstract}
We give an integral variational characterization for the speed of fronts
of the nonlinear diffusion equation $u_t = u_{xx} + f(u)$ with
$f(0)=f(1)=0$, and $f>0$ in $(0,1)$, which permits, in principle,
the  calculation of  the exact speed for arbitrary $f$.
\end{abstract}

\newpage

\section{Introduction}

The problem of the asymptotic speed of propagation of the
interface between an unstable and stable state has received much
attention in connection with different problems of population
growth, chemical reactions, pattern formation and others. We refer to
\cite{CH93} for a recent review and references. The
best understood of such problems is that of the nonlinear
reaction diffusion equation
$$ u_t = u_{xx} + f(u) \eqno(1a) $$
with
$$ f(0) = f(1) = 0,\,\, f'(0) > 0 \qquad {\rm and}\quad
f>0 \qquad {\rm in}\quad (0,1) \eqno(1b)
$$
for which Aronson
and Weinberger [AW] \cite{AW78} have shown that any positive
sufficiently localized initial condition $u(x,0)$ evolves into a
front that joins the stable state $u=1$ to $u=0$. The asymptotic
speed at which the front propagates is the minimal speed $c^*$
for which there is a monotonic front joining $u=1$ to $u=0$.
Moreover they show that the selected speed is bounded above and
below by
$$ 2\sqrt{f'(0)} \le c^* < 2 \sup \{\sqrt{f(u)\over u} \bigm| u \in
(0,1) \}  \eqno(2)
$$
and that the asymptotic selected front approaches
the fixed point $u=0$  exponentially with slope
$$
 m= -{1\over2} (c^* + \sqrt{c^{*2}-4f'(0)}). \eqno(3)
$$
 The lower bound
$c_L = 2\sqrt{f'(0)}$ is that predicted from a physical
argument, the linear marginal stability hypothesis \cite{KPP37}.
For concave functions $f$, the upper and lower bounds coincide
and the speed is exactly the linear value. However, the
asymptotic speed of propagation can still be the linear value
even when the upper and lower bounds do not coincide as explicit
examples and a variational characterization \cite{HR75} which
provides improved upper bounds show. We have recently obtained
an improved lower bound \cite{BD94} on the speed of the front
that enables one to decide when the selected speed is greater than the linear
value case which is referred to as nonlinear marginal stability selection.
There have been several reformulations of Aronson's and
Weinberger's rigorous results for the nonlinear diffusion
equation aiming to their heuristic extension to other higher order and
pattern forming equations \cite{DL83,BBDKL85,VS89,PCGO94}.
 None of these approaches however
provide the means to calculate a priori the velocity of the fronts.

The purpose of the present work is to extend our previous result  \cite{BD94}
to show an integral variational characterization of the speed of the
fronts of equation (1) which enables, in principle,  its exact calculation for
arbitrary $f$. Our main result is the following

\begin{Theorem} [Variational Characterization of $c^*$]
Let $f \in C^1(0,1)$ with $f(0)=f(1)=0$, $f'(0)>0$  and $f(u)>0$ for $u \in
(0,1)$. Then
$$
c^{*} = \max \{ 2 \sqrt{f'(0)}, J \},
$$
where
$$
J = \sup \{I(g) \bigm| g \in {\cal D} \}.
$$
Here,
$$
I(g) = 2 {\int_0^1 \sqrt{f g h} du \over \int_0^1 g du} \eqno(5)
$$
and
$\cal D$ is the space of functions in $C^1(0,1)$ such that
$g \ge 0$, $h \equiv - g' > 0$ in $(0,1)$, $g(1) = 0$, and
$\int_0^1 g(u)\, du  < \infty$.
Moreover, if $c^{*} \neq 2 \sqrt{f'(0)}$, $J$ is attained at some
$\hat g \in \cal D$, and $\hat g$ is unique up to a multiplicative
constant.
\end{Theorem}

In Section 2 we prove Theorem 1, and in Section 3 an example is given.

\section{Proof of the Variational Characterization}

We are interested in the calculation of the minimal speed for which
equation (1) has a monotonic travelling front $u(x,t) = q (z)$ with
$z =x - ct$
joining $u=1$ to $u =0$.
  Since the selected speed corresponds to that of a decreasing monotonic
front, it is convenient to work in phase space.
 Calling
$p(q) = - dq/dz$, where the minus sign is included so that $p$ is positive, we
find that the monotonic fronts are  solutions of
$$
p(q)\, {dp\over dq} - c\, p(q) + f(q) = 0, \eqno(7)
$$
with
$$
p(0) = 0, \qquad p(1) = 0, \qquad p > 0 \quad  {\rm in}\quad (0,1). \eqno(8)
$$
As shown by Aronson and Weinberger \cite{AW78}, the asymptotic speed
of propagation of fronts of equation (1), $c^*$, is the minimum value
$c$ for which there is a solution of (7) and (8). Aronson and
Weinberger have proved that there is a (unique) $p$ satisfying (7)
and (8) for $c=c^*$ (see \cite{AW78}, Section 4). Moreover, the
solution $p$ is such that $p(q) \sim \vert m \vert q$ near $q=0$,
where $\vert m \vert$ is the largest root of the equation
$$
x^2 - c^* x + f'(0) = 0,    \eqno(9)
$$
i.e.,
$$
\vert m \vert = {1\over2} (c^* + \sqrt{c^{*2}-4f'(0)}).
$$
We find it convenient to introduce the parameter $\lambda$ defined as
$\lambda = c^*/ \vert m \vert$. In terms of $\lambda$ one can write
$$
c^* = \lambda \sqrt{{f'(0)\over \lambda -1 }}\qquad  {\rm and}\qquad
\vert m \vert = \sqrt{{f'(0)\over \lambda -1}}. \eqno(10)
$$
It is straightforward to verify that whenever $1< \lambda < 2$  the
value of $\vert m \vert$ given by (10)  corresponds to the largest
root of (9) and therefore to the asymptotic slope at the origin of
the selected front \cite{BD94A}. At
$\lambda =2$ the speed $c^*$ attains the linear value $c_L$.

\noindent
{\it Proof of Theorem 1.}
The proof of the Theorem is done in two steps. First we show that
$$
c^{*} \ge  \max \{ 2 \sqrt{f'(0)}, J \},  \eqno(11)
$$
and then we show that either $c^{*}=2 \sqrt{f'(0)}$ or $c^{*}=J$. From
the results of Aronson and Weinberger [AW] it follows that
$c^{*} \ge 2 \sqrt{f'(0)}$ (see equation (2) above). Thus, to prove (11)
we need only to show that
$$
c^{*} \ge I (g)
$$
for all $g \in {\cal D}$. This latter fact has been proven by us in
\cite{BD94}.  We repeat here the argument for completeness. Let $g$ be any
function in $\cal D$.
Multiplying equation (7) by $g/p$ and integrating with respect to $q$ we find
after integrating by parts,
$$
c^* = {\int_0^1 \left( h\, p + {f(q)\over p}\, g \right) dq\over
\int_0^1 g(q) dq}
\eqno(12)
$$
However since  $p,\,h,\, f$ and $g$ are positive,
for every fixed $q$
$$
h\,p + {f(q)\, g\over p} \ge 2 \, \sqrt{f\, g\, h} \eqno(13)
$$
hence
$$
c^*\, \ge 2\, {{\int_0^1 \sqrt{ f\, g\, h}\, dq}\over{\int_0^1
 g\, dq}} \eqno(14)
$$
which proves (11).

To finish the proof of the theorem notice that if $c^{*}=2 \sqrt{f'(0)}$
we are done. Therefore, let's assume $c^* \neq 2 \sqrt{f'(0)}$ (in fact,
$c^* >  2 \sqrt{f'(0)}$). We will show that $c^*=J$ and that there
exists $\hat g \in {\cal D}$ such that $c^*=I(\hat g)$.

Let $p(q)$ be the  positive solution of (7) satisfying (8). The
existence of such a solution has been established by [AW]. Moreover,
$p(q) \sim \vert m \vert q$ near $q=0$.

In terms of $p$, let us define the function $v(q)$ as
$$
v(q) = \exp \{ \int_{q_0}^q \frac{c^*}{p} \, dq \}   \eqno(15)
$$
for some fixed $0<q_0<1$. Since $p$ is continuous and strictly
positive in $(0,1)$, $v$ is $C^1$ in $(0,1)$. Moreover, $v$ satisfies
$$
{ v'\over v} = {c^*\over p}. \eqno(16)
$$
Now choose
$$
\hat g = {1\over  v'}. \eqno(17)
$$
{}From (15) and (17) it follows that
$$
\hat g(q) = \frac{p(q)}{c^*} \exp  \{ \int_{q}^{q_0} \frac{c^*}{p} \, dq \}.
\eqno(18)
$$
Clearly, $\hat g(1)=0$, $\hat g(q)>0$, and since $p \in C^1(0,1)$,
$\hat g \in
C^1(0,1)$. In order to show that $\hat g \in \cal D$, we must verify that
$h \equiv - \hat g' >0$ in $(0,1)$.
It follows from (17) that
$$
\hat g' = - { v'' \over v'^2} \eqno(19a)
$$
and taking the derivative of (16) with respect to $q$ we have
$$
{ v'' \over v} - {v'^2\over v^2} = - {c\over p^2}\, p' \eqno(19b)
$$
Using equations (7) and (16) to eliminate $p'$ and $v'$ in
(19b) we find
$$
{v'' \over v} = {c^*  f\over p^3} \eqno(20)
$$
Since $v$, $f$ and $p$ are positive, $v''>0$; hence, it follows from
(19a) that  $h = - \hat g' > 0$ in $(-0,1)$. Alternatively, it follows from (7)
and (18) that
$$
h(q) = f(q) \exp  \{ \int_{q}^{q_0} \frac{c^*}{p} \, dq \} \ge 0.
\eqno(21)
$$
Finally we must show that $\int_0^1 \hat g(q) \, dq < \infty$.
As we have seen, $\hat g$ is a continuous, positive and decreasing
function in $(0,1)$. Hence, it is bounded away from the origin. Thus,
to determine whether $\int_0^1 \hat g(q) \, dq$ is finite we must study
the behavior of $\hat g$ near $q=0$. Since we know that $p \sim \vert m
\vert q$ near $0$, it follows from (16) that
$$
\frac{v'}{v} \sim \frac{c^*}{\vert m \vert} \frac{1}{q} = \frac{\lambda}{q},
$$
which in turns implies that
$$
\hat g \sim \frac{1}{\lambda} \frac{1}{q^{\lambda-1}}
$$
near $q=0$. Therefore, if $\lambda<2$ (i.e., if $c^* > c_L$), we have
$\int_0^1 \hat g(q) \, dq < \infty$ and $g \in {\cal D}$.

Having verified that $\hat g \in \cal D$ we now show that for this choice of
$\hat g$ the equality holds in (13). We must verify that
$$
h p + {f \hat g \over p} = 2\sqrt{f \hat g h}
$$
or equivalently, that
$$
h p = {f \hat g \over p}.
$$
This follows directly from the definition of $\hat g$, in fact from (19a)
$ h p \equiv - \hat g' p = v'' p/v'^2 $. Using equations (16), (20) and the
definition (17)  of $\hat g$
the result follows.  Having shown that whenever $\lambda < 2$ (i.e.,
whenever $c^* > c_L$) there
exists a function $\hat g \in {\cal D}$ for which we obtain the exact
speed we have proven (4).

\bigskip
\bigskip
\noindent
{\bf Note:} The uniqueness of the maximizer follows from the fact
that we need to choose $\hat g$ in such a way that (13) is satisfied as an
equality. Alternatively, we can prove uniqueness directly from the
concavity of $I(g)$. From the homogeneity of $I(g)$ in $g$ we can
restrict, without loss of generality, to $g$'s such that $\int_0^1
g(q) \, dq=1$. Assume $c^* >c_L$ and that there are two $g$'s in
${\cal D}$, $g_1$
and $g_2$ say, such that $I(g_1)=I(g_2)=c^*$ (and $\int g_1=\int g_2
=1$). Then, let $g= \alpha g_1 + (1-\alpha) g_2$, whith $0<\alpha<1$.
Clearly, $\int g=1$, and $h \equiv -g'=  \alpha h_1 + (1-\alpha) h_2
> 0$. Therefore,
$$
I(g) = \int_0^1 \sqrt{f(\alpha g_1 + (1-\alpha) g_2) (\alpha h_1 +
(1-\alpha)h_2)} \, dq
$$
However, by the Cauchy--Schwarz inequality,
$$
\sqrt{(\alpha g_1 + (1-\alpha) g_2)} \sqrt{(\alpha h_1 +
(1-\alpha)h_2} \ge \alpha \sqrt{h_1 g_1} + (1-\alpha)\sqrt{h_2 g_2}
$$
with equality if and only if $g_1=\beta h_1$ and $g_2=\beta h_2$.
Thus,
$$
c^* \ge I(g) \ge \alpha I(g_1) + (1-\alpha) I(g_2) = c^*
$$
which implies $h_1/g_1=h_2/g_2$, and therefore $g_1\equiv g_2$.

\section{Example}

In this section we illustrate the results by applying it to the
exactly solvable case $f(u) = u (1 - u) (1 + a u)$ for which it is known
that
$$
c^* = \left\{ \begin{array}{ll}
              \sqrt{2\over a} + \sqrt{a\over 2}
             & \mbox{if $a > 2$} \\
             2\sqrt{f'(0)}=2 &\mbox{if $a< 2$}
           \end{array} \right.
$$
The idea is to combine the variational characterization contained in
Theorem 1, which given a trial function yields lower bounds for
$c^*$, with the variational characterization of Hadeler and Rothe
\cite{HR75}, which provides upper bounds on  $c^*$.

For a given $a>0$, let us take
$$
g(q) = {(1 - q)^{\lambda +1}\over q^{\lambda -1} } \qquad \mbox{with
 $\lambda = 1 + {2\over a}$}.
$$
Then
$$
h = -g' = { (1 - q)^\lambda\over q^\lambda} (\lambda -1) (1 + a q)
$$
and \cite{AStables}
$$
\int_0^1 g(q)\, dq = {\Gamma (\lambda + 2) \Gamma (2-\lambda)\over
\Gamma (4) } \qquad {\rm if}\quad \lambda < 2,
$$
i.e., if $a>2$.

We obtain
$$
\int_0^1 \sqrt{f g h}\, dq = \sqrt{\lambda -1 } \left( \int_0^1
{(1-q)^{\lambda -1}\over q^{\lambda -1}} dq + a \int_0^1 (1 -
q)^{\lambda +1} q^{2-\lambda} dq \right)
$$
and therefore
$$
I(g) = 2\sqrt{\lambda -1} \left( 1 + a {\Gamma (3-\lambda)\over
\Gamma (2-\lambda)} {\Gamma (4)\over \Gamma (5)} \right)
$$
Using the definition of $\lambda$ and $\Gamma (z+1) = z \Gamma (z)$
we obtain
$$
I(g) = \sqrt{2\over a} + \sqrt{a\over 2}
$$
for $a>2$. Therefore, from Theorem 1 we have that
$$
c^* \ge \left\{ \begin{array}{ll}
              \sqrt{2\over a} + \sqrt{a\over 2}
             & \mbox{if $a > 2$} \\
             2\sqrt{f'(0)}=2 &\mbox{if $a< 2$}
           \end{array} \right.
$$
To prove the reversed inequality we use Hadeler and Rothe's
variational characterization \cite{HR75}:
$$
c^* \le \sup_u \{ \dot \rho (u)+ \frac{f(u)}{\rho(u)} \}
$$
for any $\rho \in C^1(0,1)$ such that $\rho(0)=0$ and $\rho(u)>0$ in
$(0,1)$.

For $a \ge 2$ choose $\rho(u)=\sqrt{a/2} \, \,  u(1-u)$. Then
$\dot \rho + (f/\rho) = \sqrt{a/2} + \sqrt{2/a}$.
For $a \le 2$ choose $\rho=u(1-u)$. Hence, $\dot \rho + (f/\rho) =
2+(a-2)u$ and $\sup_{0\le u \le 1}({\dot \rho + (f/\rho)}) =2$.
Therefore, from Hadeler's and Rothe's bound we get
$$
c^* \le \left\{ \begin{array}{ll}
              \sqrt{2\over a} + \sqrt{a\over 2}
             & \mbox{if $a > 2$} \\
             2\sqrt{f'(0)}=2 &\mbox{if $a< 2$}
           \end{array} \right.
$$
which combined with our lower bound gives the desired result.

\section{Conclusion}

We have given a variational characterization of the minimal speed for
which the nonlinear diffusion equation has monotonic fronts. As [AW]
have shown this is the asymptotic speed of propagation of
a sufficiently localized positive initial condition $u(x,0)$.

The variational principle we have derived here can also be used to study
the dependence of $c$ on the parameters of $f$. Monotonicity properties
can be immediatly derived. Derivatives of $c$ with respect to
parameters of $f$ can be obtained using the Feynman--Hellmann formula.

\section{Acknowledgments}

 This work was partially
supported by Fondecyt project 193-0559.

\end{document}